%% file: draft_final.tex
\documentclass[twocolumn,showpacs,preprintnumbers,amsmath,amssymb,prl,superscriptaddress]{revtex4}

\usepackage{graphicx} % Include figure files
\usepackage{dcolumn}  % Align table columns on decimal point
\usepackage{colordvi}
\usepackage{color}
\graphicspath{{ps}}

\begin{document}

\preprint{\vbox{ \hbox{   }
    \hbox{}
    \hbox{}
    \hbox{}
%    \hbox{hep-ex nnnn, if available}
}}

\title { \quad\\[0.5cm] Measurement of $B(D_s^+\to\mu^+\nu_\mu)$}

\input{author_list.tex}

%\collaboration{The Belle Collaboration}

\date{\today}% It is always \today, today,
             %  but any date may be explicitly specified

\begin{abstract}

We present a measurement of the branching fraction
$B(D_s^+\to\mu^+\nu_\mu)$ 
using a 548~fb$^{-1}$ 
data sample collected by the Belle 
experiment at the KEKB $e^+e^-$ collider. 
The $D_s$ momentum is
determined by reconstruction of the system recoiling against 
$DK\gamma X$ in events of the type $e^+e^-\to D_s^\ast DKX, 
D_s^\ast \to D_s\gamma$, where $X$ represents 
additional pions or photons from fragmentation. The full
reconstruction method provides high 
resolution in the neutrino momentum and thus good background
separation, equivalent to that 
reached by experiments at the tau-charm factories. We obtain the 
branching fraction $B(D_s^+\to\mu^+\nu_\mu) = (6.44 \pm  0.76({\rm stat})
\pm 0.57({\rm syst})) \cdot  10^{-3}$, 
implying a $D_s$ decay constant of $f_{D_s} = (275 \pm  16({\rm stat}) \pm
12({\rm syst}))$~MeV.
\end{abstract}
% PACS, the Physics and Astronomy
\maketitle

One of the important goals of particle physics is the precise
measurement and understanding of the 
Cabibbo-Kobayashi-Maskawa (CKM) matrix elements, 
fundamental 
parameters of the
Standard Model (SM). To interpret precise experimental results on decays of $B$
mesons 
in terms of the CKM matrix elements, theoretical calculations of 
form factors and decay 
constants (usually based on lattice gauge theory, 
see e.g. \cite{ref1}) are needed. Decays of charmed hadrons in turn
enable 
tests of the predictions for analogous quantities
in the charm sector.  
Measurements of
charmed meson decay rates with an accuracy that matches the
precision of theoretical calculations is thus necessary for checks and
further tuning of theoretical methods. 

The purely leptonic decay
$D_s^+\to{\cal{\ell}}^+\nu_{\cal{\ell}}$ 
(the charge-conjugate mode is implied throughout this paper) is
theoretically rather clean; 
in the SM, the decay is mediated by a single virtual
$W^\pm$ boson. The decay rate is given by
\begin{equation}
\Gamma(D_s^+\to
      {\cal{\ell}}^+\nu_{\cal{\ell}})=\frac{G_F^2}{8\pi}f_{D_s}^2
      m_{\cal{\ell}}^2
      m_{D_s}\bigl(1-\frac{m_{\cal{\ell}}^2}{m_{D_s}^2}\bigr)^2 \vert
      V_{cs}\vert^2~~,
\label{eq1}
\end{equation}
where $G_F$ is the Fermi coupling constant, $m_{\cal{\ell}}$ and
$m_{D_s}$ are the masses of the lepton and of the $D_s$ meson,
respectively. 
$V_{cs}$ is the corresponding CKM matrix element, while all effects of
the strong interaction are 
accounted for by the decay constant $f_{D_s}$. 
While the decay rate is tiny 
for electrons due to the strong helicity suppression and since the
detection of $\tau$'s involve 
additional neutrinos, the muon mode is experimentally the cleanest and
the most accessible one. Decays with electrons can be used to study
the backgrounds. 

The analysis described in this paper uses data
from the Belle experiment \cite{ref2} at the KEKB collider \cite{ref3}
corresponding to 
548~fb$^{-1}$. We study the decay $D_s^+\to\mu^+\nu_\mu$ using the
full-reconstruction recoil method 
first established in the study of semileptonic $D$ mesons
\cite{ref4}. Similar analyses have also been performed by the CLEO-c
\cite{ref5} and BaBar \cite{ref6} experiments. 

The Belle detector is a
large-solid-angle magnetic 
spectrometer that consists of a silicon vertex detector (SVD), a  
50-layer central drift chamber (CDC), 
an array of aerogel threshold Cherenkov counters (ACC), 
a barrel-like arrangement of time-of-flight scintillation counters
(TOF), and an electromagnetic calorimeter 
comprised of CsI(Tl) crystals (ECL) located inside a superconducting
solenoid coil that provides a 1.5~T magnetic field. An iron
flux-return located outside of the coil is 
instrumented to detect $K^0_L$ mesons and to identify muons (KLM). The
detector is described in 
detail elsewhere \cite{ref2}. Two inner detector configurations were
used. A 2.0~cm beampipe and a 3-layer silicon vertex detector were
used for the first sample of 156~fb$^{-1}$, while a 1.5~cm beampipe, a
4-layer silicon detector and a small-cell inner drift chamber were
used to record the remaining 
392~fb$^{-1}$ \cite{ref7}. 

This analysis uses events of the type
$e^+e^-\to D_s^\ast D^{\pm,0} K^{\pm,0}X$, where $X$ can be any number
of additional pions from 
fragmentation, and up to one photon \cite{ref8}. The {\it tag side}
consists of a $D$ and a 
$K$ meson (in any charge combination) while the {\it signal side} is a
$D_s^\ast$ meson decaying to $D_s\gamma$. 
Reconstructing the tag side,
and allowing any possible set of particles in $X$, the signal side is
identified by reconstruction of the recoil mass, using the known beam
momentum and four-momentum conservation. 

Tracks are detected with the CDC and the SVD. They are required to
have at least one associated hit in the SVD and an impact parameter
with respect to the interaction point of less than 2~cm in the radial
direction and less than 4~cm in the beam direction. 
Tracks are also
required to have momenta in the 
laboratory frame greater than 100~MeV/$c$. A likelihood ratio for a
given track to be a kaon or pion, ${\cal{L}}(K,\pi)$, is obtained by
utilizing specific 
ionization energy loss measurements in the CDC, light yield
measurements from the ACC, and time-of-flight information from the TOF
\cite{ref9}. We require 
 ${\cal{L}}(K,\pi)>0.5$ for kaon candidates. 
The momentum of the 
lepton candidates is required to be larger than 500~MeV/$c$. 
For electron identification we use position, cluster
energy, shower shape in the ECL, 
combined with track momentum and $dE/dx$ measurements in the CDC and
hits in the ACC. For muon 
identification, we extrapolate the CDC track to the KLM and compare
the measured range and 
transverse deviation in the KLM with the expected values. Photons are
required to have energies in 
the laboratory frame of at least 50 - 150~MeV, depending on the
detecting part of the ECL. 
Neutral pion candidates are reconstructed using photon pairs with
invariant mass within $\pm$10~MeV/$c^2$ 
of the nominal $\pi^0$ mass. Neutral kaon candidates are reconstructed
using charged pion pairs with invariant mass 
within $\pm$30~MeV/$c^2$ of the nominal $K^0$ mass. 

Charged and neutral tag-side $D$ mesons are re-constructed in
$D\to Kn\pi$ decays with $n = 1,2,3$ (total branching fraction $\approx
25\%$). Mass windows were 
optimized for each channel separately, and a 
mass-constrained vertex fit (requiring a confidence level 
greater than 0.1\%) is applied to the $D$ meson to improve the momentum
resolution. 
$D_s^\ast$-candidates are not directly reconstructed: 
we construct the mass of the system recoiling against $DKX$, using the 
known beam momentum, and require a value within $\pm~150$~MeV/$c^2$ 
of the nominal $D_s^\ast$ mass \cite{ref10}. 
A
recoil mass $M_{\rm rec}(Y)$ is defined as the magnitude of 
the four-momentum $p_{\rm beams} -  p_Y$, for an arbitrary set of
reconstructed particles $Y$. $p_{\rm beams}$ is the momentum of the
initial $e^+e^-$ system. 
Since at this point in the
reconstruction $X$ can be any set of remaining pions and photons,
there is usually a large number of combinatorial possibilities. It is
reduced by requiring the presence of a photon that is consistent with
the decay $D_s^\ast \to D_s\gamma$, where the $D_s$ mass lies within 
$\pm$150~MeV/$c^2$ of its nominal mass \cite{ref10}. Further selection
criteria are applied on the 
momenta of particles in the $e^+e^-$ rest frame; for the primary $K$
meson the momentum should be smaller than 2~GeV/$c$, for the $D$ meson larger than
2~GeV/$c$ and for the $D_s$ meson larger than 3~GeV/$c$. 
The energy of the photon from $D_s^\ast \to D_s\gamma$ in
the lab frame is required to be larger than 150~MeV,
irrespective of its polar angle. To further improve the recoil
momentum resolution, 
inverse \cite{ref11} mass-constrained vertex fits are then performed
for the $D_s^\ast$ and $D_s$, requiring a confidence level greater
than 1\%. 
After implying these selection criteria, the average number of
combinatorial reconstruction possibilities 
is approximately 2 per event. The sample is further divided into a right-
(RS) and wrong-sign (WS) part. If the primary $K$ meson is charged,
    both it and the $D$ meson are required to have opposite flavor
    (strangeness or charm respectively) to the  $D_s^\ast$, to be
    counted in the right-sign sample; all other combinations are
    wrong-sign. If the primary $K$ meson is a $K^0_S$, the assignment is
    based on the relative flavor of the $D$ and $D_s^\ast$ mesons
    alone. 
The flavor of the $D_s^\ast$ is 
fixed by the total charge of the $X$, 
assuming overall charge 
conservation for the event.   

\begin{figure}[t]
  \includegraphics[width=8.57cm]{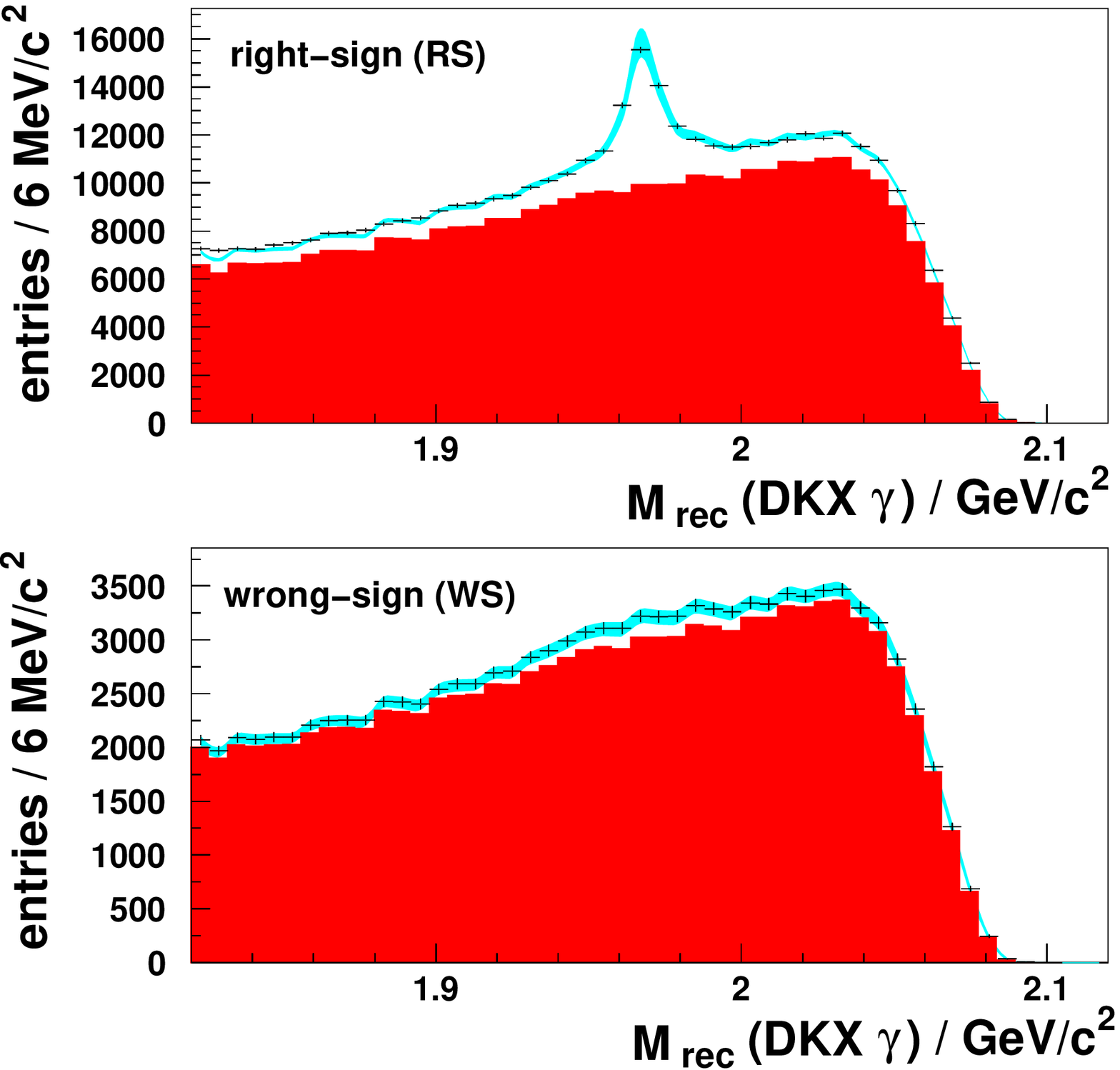}
\includegraphics[width=8.57cm]{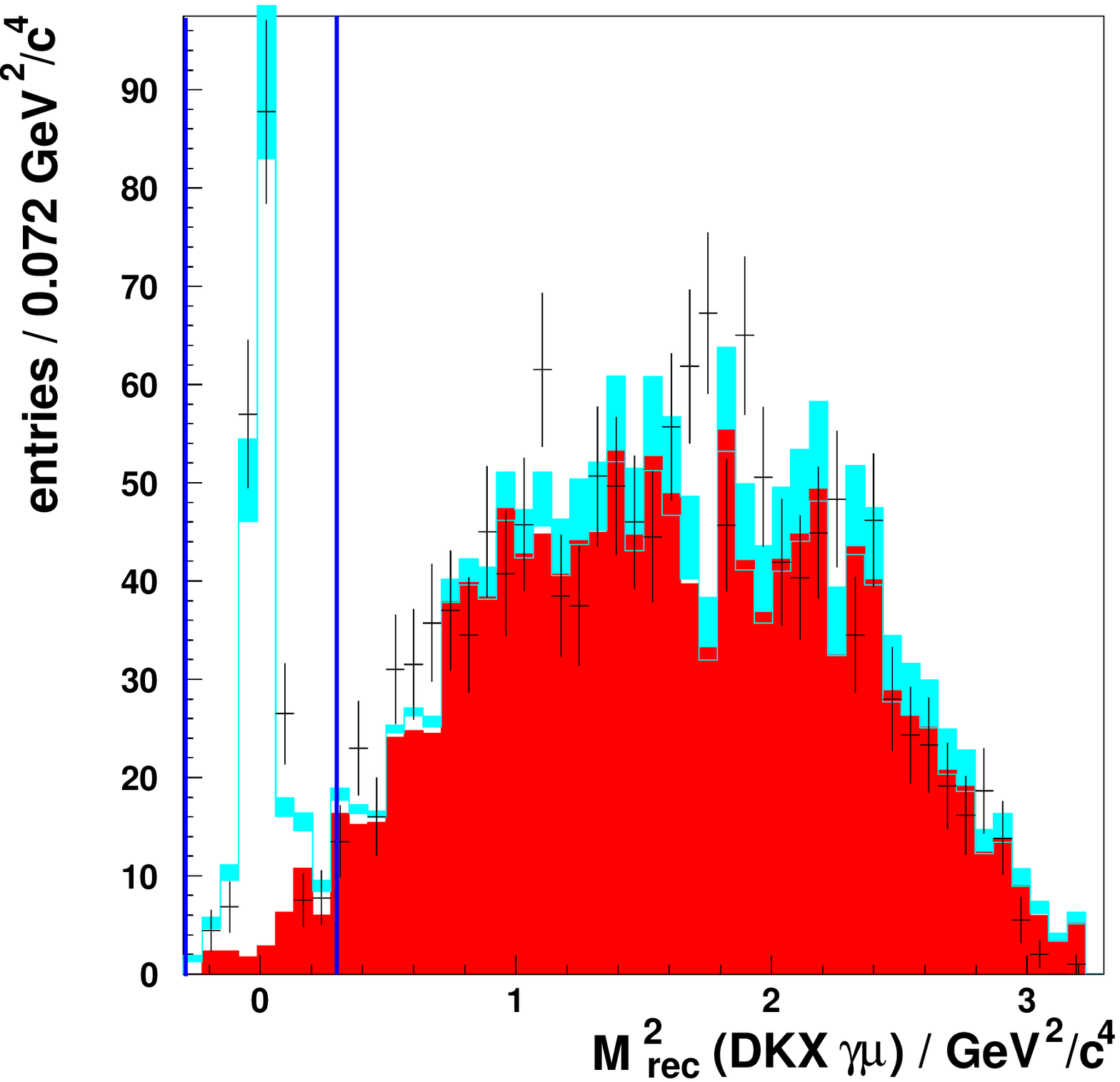}
  \caption{Top: recoil mass spectrum for $D_s$-tags (for both RS
and WS samples). Bottom: spectrum of missing mass 
squared for $D_s^+\to\mu^+\nu_\mu$ candidates for the selected
data. 
Error bars represent the statistical errors. The dark-shaded 
areas show the fitted 
background, the light-shaded bands show the fit with systematic
uncertainties. The vertical lines indicate 
the signal regions.}
  \label{fig1}
\end{figure}

Within this sample of tagged inclusive $D_s$ decays (named $D_s$-tags
in the following), decays of the type $D_s\to
\mu\nu_\mu$ are selected by requiring 
another charged track that is identified as a muon and has the same
charge as the $D_s$ candidate. No additional charged particles are
allowed in the event. Remaining photons not used in the 
described reconstruction are allowed only if their total energy is smaller
than $1.0/m$~GeV, where $m$ is the number of such particles. 
After these selections, in almost 
all cases only one combinatorial reconstruction possibility remains. 
Figure \ref{fig1} shows the mass spectra of $M_{\rm
  rec}(DKX\gamma)$ 
(corresponding to the candidate $D_s$
mass) and of $M_{\rm rec}(DKX\gamma\mu)$ (corresponding to the neutrino candidate mass).

We define $n_X$ as the number of {\it primary} particles in the event,
where primary means that the 
particle is not a daughter of any particle reconstructed in the
event. The minimal value for $n_X$ is three corresponding to a
$e^+e^-\to D_s^\ast D K$ event without any further particles from
fragmentation. The upper limit for $n_X$ 
is determined by the reconstruction efficiency; Monte Carlo (MC)
simulation shows that the number of reconstructed signal events is
negligible for $n_X > 10$. As the 
efficiency very sensitively depends on $n_X$, it is crucial to use MC
simulation that correctly reflects the $n_X$ distribution observed in
the data. Unfortunately, the details of fragmentation processes are
not very well understood, and standard MC events show notable
differences compared to the data. 
Furthermore, the true (generated) $n^T_X$ value differs from the
reconstructed $n^R_X$, as particles 
can be lost or wrongly assigned. Thus the measured (reconstructed)
$n^R_X$ distribution has to be 
deconvoluted so that the analysis can be done in bins of $n^T_X$ to
avoid bias in the results. 

To extract the number of $D_s$-tags as a function of $n^T_X$ in data, two
dimensional simulated distributions in $n^R_X$ (ranging from 3 to 8) and the
recoil mass $M_{\rm rec}(DKX\gamma)$ are fitted to the
RS and WS data distributions. 
The signal shapes for
different values of $n^T_X$ (ranging from 3 to 9 \cite{ref12}) of the
signal are modeled with 
generic MC simulation \cite{ref13}, which has been filtered at the
generator level for events of the type $e^+e^-\to D_s^\ast D KX$. The
weights of these components, 
$w^{D_s}_i, i = 3,... 8$, are free parameters in the fit to
the data. As a model for the background in 
the RS sample, the WS data sample is used. The normalization constants
between WS and RS (which vary with $n^R_X$) are another six fit
parameters. Since the WS sample contains some signal ($\approx 10\%$
of the RS signal), these signal components for different $n^T_X$ values
are also included 
in the fit as
independent parameters.  
As a 
cross-check, the fit has also been performed using generic MC RS-sample
backgrounds, which gives a negligible change in the results. A further
cross-check involved the random 
division of the MC sample into two halves, using the shapes of the
first half to fit the signal in 
the second. The resulting weights as function of $n^T_X$ 
fit to a constant of $0.990 \pm
0.046$, which agrees well 
with the expectation of 1. The total number of reconstructed
$D_s$-tags in data is calculated as 
\begin{equation}
N_{D_s}^{{\rm rec}}=\sum_{i=3}^8w_i^{D_s}N_{D_s}^{{\rm MC}, i}~~,
\label{eq2}
\end{equation}
where $N_{D_s}^{{\rm MC}, i}$ represents the total number of reconstructed
filtered MC events that were 
generated with $n^T_X=i$ (regardless of the reconstructed $n^R_X$) and
$w_i^{D_s}$ the fitted weight of this component. 

To fit the number of $D_s\to\mu\nu_\mu$ events as a function of $n^T_X$,
two-dimensional histograms in $n^R_X$ 
and the recoil mass $M_{\rm rec}(DKX\gamma\mu)$ are used. The
shape of the signal is modeled with 
signal MC distribution. As MC studies show, the background under the $\mu\nu_\mu$
signal peak consists 
primarily of non-$D_s$ decays ($\approx 18\%$ of signal), leptonic
$\tau$ decays (where the $\tau$ decays to a muon and two neutrinos,
$\approx 7\%$) and semileptonic $D_s$ decays (where the additional
hadrons have low momenta and 
remain undetected, $\approx 3.6\%$). Hadronic $D_s$ decays (with one
hadron misidentified as a muon) are a rather small background component
($\le 2\%$ of signal). Except for hadronic decays, which are
negligible, all backgrounds are common 
to the $e\nu_e$ mode, which is suppressed by a factor of
${\cal{O}}(10^5)$. Thus, the $e\nu_e$ sample provides a good model of
the $\mu\nu_\mu$ background that has to be corrected only for kinematical
and efficiency differences. Including this corrected shape in the fit,
the total number of fitted 
$\mu\nu_\mu$ events in data is given by 
\begin{equation}
N_{\mu\nu}^{{\rm rec}}=\sum_{i=3}^8w_i^{\mu\nu}N_{\mu\nu}^{{\rm MC}, i}~~,
\label{eq3}
\end{equation}
where $N_{\mu\nu}^{{\rm MC}, i}$ represents the total number of
reconstructed signal MC events that were 
generated in the $i$-th bin of $n^T_X$ (regardless of the
reconstructed $n^R_X$ ) and $w_i^{\mu\nu}$ is the fitted weight of
this component. 

The numerical result for $N_{D_s}^{{\rm rec}}$ is $32100
\pm  870({\rm stat}) \pm 1210({\rm syst})$, that for
$N_{\mu\nu}^{{\rm rec}}$ is $169 \pm  16({\rm stat}) \pm  8({\rm
  syst})$. The statistical errors reflect the finite number of data
signal candidates. The systematic errors are due to the limited statistics of
WS data and MC signal and background samples. The errors were estimated
by varying the bin contents of data
and MC distributions and repeating the fits. By this procedure 
the non-negligible correlations among the fitted
weights were taken into account. 

As the
branching fraction of $D_s\to\mu\nu_\mu$ used for the generation of
MC events is known, the branching fraction in data can be
determined using the following 
formula:
\begin{equation}
{\cal{B}}(D_s\to\mu\nu_\mu)=\frac{N_{\mu\nu}^{{\rm
      rec}}}{\overline{\epsilon}_{\mu\nu}{N_{D_s}^{{\rm rec}}}}=
\frac{N_{\mu\nu}^{{\rm rec}}}{N_{\mu\nu}^{{\rm MC},{\rm
      rec}}}{\cal{B}}_{{\rm MC}}(D_s\to\mu\nu_\mu)~~,
\label{eq4}
\end{equation}
where ${\cal{B}}_{{\rm MC}}(D_s\to\mu\nu_\mu)= 0.51\%$ and
$N_{\mu\nu}^{{\rm MC},{\rm rec}}$ is the number of reconstructed $\mu\nu_\mu$ events
in MC simulation, weighted according to the fit to data, i.e.
\begin{equation}
N_{\mu\nu}^{{\rm MC},{\rm rec}}=\sum_{i=3}^8w_i^{D_s}N_{\mu\nu}^{{\rm MC}, i}.
\label{eq5}
\end{equation}
The average efficiency for the
reconstruction of $D_s\to\mu\nu_\mu$ decays,
$\overline{\epsilon}_{\mu\nu}$, is not needed explicitly
for the computation of the branching fraction \cite{ref15}. The final
result is:
\begin{equation}
{\cal{B}}(D_s\to\mu\nu_\mu)\cdot 10^3=6.44 \pm  0.76({\rm stat}) \pm
0.57({\rm syst}). 
\label{eq6}
\end{equation}
The quoted statistical error reflects the statistical uncertainty of
the fitted weights $w_i^{D_s}$ and $w_i^{\mu\nu}$, including 
their correlations. 
The systematic error combines the
contributions due to the statistical 
uncertainties of data and MC background samples (0.29), the
statistical uncertainty of the signal MC distribution (0.41), muon tracking
and identification efficiency (0.18) and possible differences in relative rates
of individual $D_s$ decay modes 
between MC simulation and data (0.19). Since the branching fraction is
determined relative to the number 
of $D_s$-tags, the systematic errors in the reconstruction of the tag
side cancel. Differences in the 
neutrino peak resolution between data and simulation have been found
to have a negligible effect on the systematic error. 

\begin{figure}[t]
  \includegraphics[width=8.57cm]{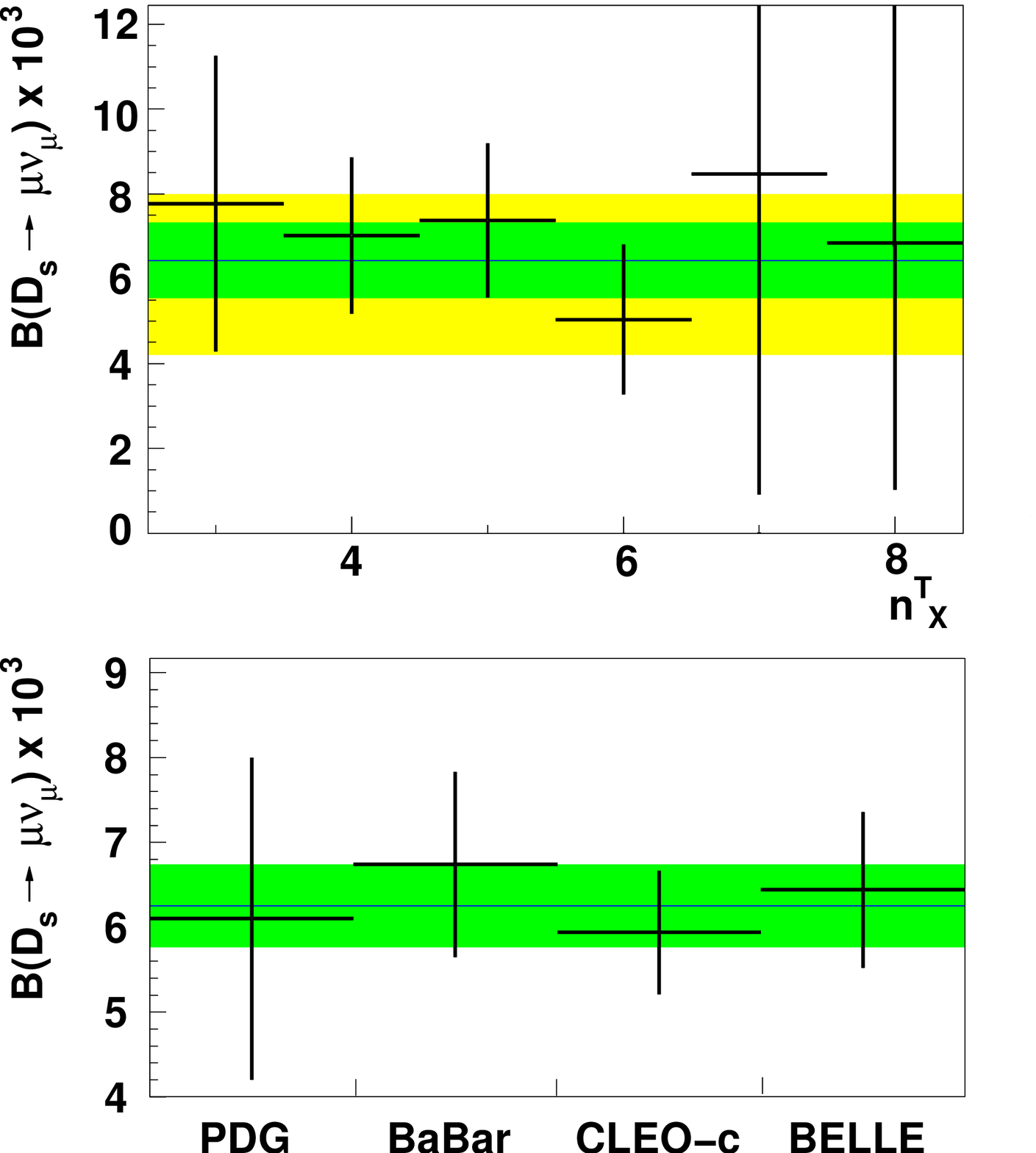}
  \caption{Top: ${\cal{B}}(D_s\to\mu\nu_\mu)$ as a function of
    $n^T_X$; the final result is shown as the
dark-shaded region. 
For comparison, the PDG value and its error is shown as the light-shaded
region in the background. Bottom: our result compared with PDG \cite{ref10} and recent BaBar
    \cite{ref6} and CLEO-c \cite{ref5} measurements not 
yet included in the PDG 2006 compilation; the dark-shaded region shows
the weighted average of all measurements.}
  \label{fig2}
\end{figure}

Figure \ref{fig2} (top) shows the branching fraction determined in bins of
$n^T_X$. The result is stable within errors in $n^T_X$; note that the
errors shown for the $n^T_X$ bins are correlated. As a cross check,
also the branching fraction in a 
limited range $n^T_X\le 6$ has been determined as $(6.54\pm 0.76({\rm
  stat})\pm 0.57({\rm syst}))\cdot 10^{-3}$, which agrees well with
the result given above. 
Figure \ref{fig2} (bottom) shows our result in comparison with the PDG
\cite{ref10} value and recent results from other experiments
\cite{ref5,ref6}. 

In conclusion, we have studied events of the type $e^+e^-\to D_s^\ast
D^{\pm,0} K^{\pm,0}X, D_s^\ast \to D_s\gamma$ with $X=n\pi(\gamma)$
where the $D_s$ is identified in 
the recoil of the remainder of the event. Normalizing to this sample
of $D_s$-tags, the branching fraction of 
$D_s\to\mu\nu_\mu$ was measured to be $(6.44 \pm  0.76({\rm stat}) \pm
0.57({\rm syst}))\cdot 10^{-3}$, which is in good agreement with the
current PDG value of $(6.1 \pm  1.9) \cdot  10^{-3}$ \cite{ref10} and
also compatible with recent results from BaBar $(6.74 \pm  1.09) \cdot
10^{-3}$ \cite{ref6} and CLEO-c $(5.94 \pm  0.73) \cdot  10^{-3}$
\cite{ref5}. Finally we obtain the 
decay constant $f_{D_s}$, using Eqn. (\ref{eq1}) (with $\vert
V_{cs}\vert = 0.9730$ \cite{ref10})
\begin{equation}
f_{D_s} = (275 \pm  16({\rm stat}) \pm  12({\rm syst}))~{\rm MeV}.
\label{eq7}
\end{equation}

A simple average of the decay constants following from the cited
measurements has an uncertainty of around 10~MeV. Recently an LQCD
calculation of significantly improved precision was performed,
with the result $f_{D_s}=(241\pm 3)$~MeV \cite{ref14}. This 
value is somewhat lower than the experimental average and the
comparison with the experimental results may point to some
inconsistency between the two. 
More precise measurements are needed for a firm 
comparison 
and will become possible in the near future
at both $B$ and tau-charm factories.

We thank the KEKB group for excellent operation of the accelerator,
the KEK cryogenics group for efficient solenoid operations, and the
KEK computer group and the NII for valuable 
computing and Super-SINET network support. We acknowledge support
from MEXT and JSPS (Japan); ARC and DEST 
(Australia); NSFC and KIP of CAS (contract No. 10575109 and
IHEP-U-503, China); DST (India); the BK21 program of MOEHRD, and the
CHEP SRC and BR (grant No. 
R01-2005-000-10089-0) programs of KOSEF (Korea); KBN (contract
No. 2P03B 01324, Poland); MES and RFAAE (Russia); 
ARRS(Slovenia); SNSF (Switzerland); NSC and MOE (Taiwan); and DOE (USA).

\end{document}

%% file: author_list.tex
%%% Paper:    Ds -> mu nu
%%% Journal:  Physical Review Letters
%%% Contacts: L. Widhalm (widhalm@qhepu3.oeaw.ac.at)
%%% Non-responding authors or those who said NO are commented out.
%%% ====================================================================
%%% Click the RELOAD button on your web browser to see the updated file.
%%% ====================================================================
%%% Use \input{author} to insert this material into your latex file.
%%%%% Force institutions to appear in alphabetical order when typeset.
\affiliation{Budker Institute of Nuclear Physics, Novosibirsk}
\affiliation{Chiba University, Chiba}
\affiliation{University of Cincinnati, Cincinnati, Ohio 45221}
%%%\affiliation{Department of Physics, Fu Jen Catholic University, Taipei}
%%%\affiliation{Justus-Liebig-Universit\"at Gie\ss{}en, Gie\ss{}en}
\affiliation{The Graduate University for Advanced Studies, Hayama}
%%%\affiliation{Gyeongsang National University, Chinju}
\affiliation{Hanyang University, Seoul}
\affiliation{University of Hawaii, Honolulu, Hawaii 96822}
\affiliation{High Energy Accelerator Research Organization (KEK), Tsukuba}
%%%\affiliation{Hiroshima Institute of Technology, Hiroshima}
\affiliation{University of Illinois at Urbana-Champaign, Urbana, Illinois 61801}
\affiliation{Institute of High Energy Physics, Chinese Academy of Sciences, Beijing}
\affiliation{Institute of High Energy Physics, Vienna}
\affiliation{Institute of High Energy Physics, Protvino}
\affiliation{Institute for Theoretical and Experimental Physics, Moscow}
\affiliation{J. Stefan Institute, Ljubljana}
\affiliation{Kanagawa University, Yokohama}
\affiliation{Korea University, Seoul}
%%%\affiliation{Kyoto University, Kyoto}
\affiliation{Kyungpook National University, Taegu}
\affiliation{\'Ecole Polytechnique F\'ed\'erale de Lausanne (EPFL), Lausanne}
\affiliation{Faculty of Mathematics and Physics, University of Ljubljana, Ljubljana}
\affiliation{University of Maribor, Maribor}
\affiliation{University of Melbourne, School of Physics, Victoria 3010}
\affiliation{Nagoya University, Nagoya}
\affiliation{Nara Women's University, Nara}
\affiliation{National Central University, Chung-li}
\affiliation{National United University, Miao Li}
\affiliation{Department of Physics, National Taiwan University, Taipei}
\affiliation{H. Niewodniczanski Institute of Nuclear Physics, Krakow}
\affiliation{Nippon Dental University, Niigata}
\affiliation{Niigata University, Niigata}
\affiliation{University of Nova Gorica, Nova Gorica}
\affiliation{Osaka City University, Osaka}
\affiliation{Osaka University, Osaka}
\affiliation{Panjab University, Chandigarh}
%%%\affiliation{Peking University, Beijing}
%%%\affiliation{University of Pittsburgh, Pittsburgh, Pennsylvania 15260}
%%%\affiliation{Princeton University, Princeton, New Jersey 08544}
\affiliation{RIKEN BNL Research Center, Upton, New York 11973}
\affiliation{Saga University, Saga}
\affiliation{University of Science and Technology of China, Hefei}
\affiliation{Seoul National University, Seoul}
%%%\affiliation{Shinshu University, Nagano}
\affiliation{Sungkyunkwan University, Suwon}
\affiliation{University of Sydney, Sydney, New South Wales}
%%%\affiliation{Tata Institute of Fundamental Research, Mumbai}
\affiliation{Toho University, Funabashi}
\affiliation{Tohoku Gakuin University, Tagajo}
\affiliation{Tohoku University, Sendai}
\affiliation{Department of Physics, University of Tokyo, Tokyo}
\affiliation{Tokyo Institute of Technology, Tokyo}
\affiliation{Tokyo Metropolitan University, Tokyo}
\affiliation{Tokyo University of Agriculture and Technology, Tokyo}
%%%\affiliation{Toyama National College of Maritime Technology, Toyama}
\affiliation{Virginia Polytechnic Institute and State University, Blacksburg, Virginia 24061}
\affiliation{Yonsei University, Seoul}
% \author{K.~Abe}\affiliation{High Energy Accelerator Research Organization (KEK), Tsukuba} % KEK
 \author{L.~Widhalm}\affiliation{Institute of High Energy Physics, Vienna} % Vienna
   \author{I.~Adachi}\affiliation{High Energy Accelerator Research Organization (KEK), Tsukuba} % KEK
   \author{H.~Aihara}\affiliation{Department of Physics, University of Tokyo, Tokyo} % Tokyo
% \author{D.~Anipko}\affiliation{Budker Institute of Nuclear Physics, Novosibirsk} % BINP
% \author{K.~Arinstein}\affiliation{Budker Institute of Nuclear Physics, Novosibirsk} % BINP
% \author{T.~Aso}\affiliation{Toyama National College of Maritime Technology, Toyama} % Toyama
% \author{V.~Aulchenko}\affiliation{Budker Institute of Nuclear Physics, Novosibirsk} % BINP
   \author{T.~Aushev}\affiliation{\'Ecole Polytechnique F\'ed\'erale de Lausanne (EPFL), Lausanne}\affiliation{Institute for Theoretical and Experimental Physics, Moscow} % ITEP
% \author{T.~Aziz}\affiliation{Tata Institute of Fundamental Research, Mumbai} % Tata
% \author{S.~Bahinipati}\affiliation{University of Cincinnati, Cincinnati, Ohio 45221} % Cincinnati
   \author{A.~M.~Bakich}\affiliation{University of Sydney, Sydney, New South Wales} % Sydney
   \author{V.~Balagura}\affiliation{Institute for Theoretical and Experimental Physics, Moscow} % ITEP
% \author{Y.~Ban}\affiliation{Peking University, Beijing} % Peking
% \author{S.~Banerjee}\affiliation{Tata Institute of Fundamental Research, Mumbai} % Tata
   \author{E.~Barberio}\affiliation{University of Melbourne, School of Physics, Victoria 3010} % Melbourne
% \author{M.~Barbero}\affiliation{University of Hawaii, Honolulu, Hawaii 96822} % Hawaii
   \author{A.~Bay}\affiliation{\'Ecole Polytechnique F\'ed\'erale de Lausanne (EPFL), Lausanne} % Lausanne
   \author{I.~Bedny}\affiliation{Budker Institute of Nuclear Physics, Novosibirsk} % BINP
% \author{K.~Belous}\affiliation{Institute of High Energy Physics, Protvino} % Protvino
   \author{V.~Bhardwaj}\affiliation{Panjab University, Chandigarh} % Panjab
% ...
% BINP
\collaboration{The Belle Collaboration}

%% file: draft_final.bbl
\begin{thebibliography}{99}
\bibitem{ref8} It has been found that events with additional kaons or
  more than one photon have a poor 
signal/background ratio and have been therefore excluded.
\bibitem{ref11} The fit is called inverse since it uses information
  from the mother and sister particles, rather than information about
  daughter particles as is usually the case.
\bibitem{ref12} The upper limit of 9 is chosen because the bin $n^T_X
  = 9$ has some overlap with $n^R_X \le 8$. The corresponding fitted
  weight turns out to be negligible and 
is neglected in the further analysis.
\end{thebibliography}
